\documentclass[,final] {aipproc}

\layoutstyle{6x9}
\usepackage{color} 
 
\begin{document}

\title{Brain complexity born out of criticality}\footnote{in ``Physics, Computation, and the Mind - Advances and Challenges at Interfaces-".  (J. Marro, P. L. Garrido \& J. J. Torres, Eds.)  American Institute of Physics (2012, in press).}

\author{Enzo Tagliazucchi }{address={Neurology Department and Brain Imaging Center, Goethe University, Frankfurt am Main, Germany.}
}
\author{Dante R Chialvo}{address={Consejo Nacional de Investigaciones Cient\'{i}ficas y Tecnol\'{o}gicas, Buenos Aires, Argentina.}
}
\begin{abstract}
In this essay we elaborate on recent evidence demonstrating the presence of a second order phase transition in human brain dynamics  and discuss its consequences for  theoretical approaches to brain function.  We review early evidence of criticality in brain dynamics at different spatial and temporal scales, and we stress how it was necessary to unify concepts and analysis techniques across scales to introduce the adequate order and control parameters which define the transition. A discussion on the relation between structural vs. dynamical complexity exposes future steps to understand the dynamics of the connectome (structure) from which emerges the cognitome (function).
\end{abstract}
\maketitle
\section{Why criticality?}
Complexity, in simple terms, is all about how diversity and non-uniformity \cite{Bak95} arises from the uniform interaction of similar units. In all cases, the dynamics of the emergent complex behavior of the whole cannot be directly anticipated from the knowledge of the laws of motion of the isolated parts. Early forerunners of complexity science, namely statistical mechanics and condensed matter physics, have identified a peculiar scenario at which, under certain general conditions, such complexity can emerge: near the critical point of a second order phase transition.
 At this point, complexity appears as a product of the competition between ordering  and disordering collective tendencies, such that the final result is a state with a wide variety of dynamic patterns exhibiting a mixture of order and disorder.  

As argued elsewhere \cite{bak,chialvo2010} the dynamics of the human brain exhibit a large degree of concordance with those expected for a system near criticality. From the cognitive  side, brain's complexity is an almost obvious statement: the ultimate products of such complexity are, for instance, the nearly unpredictable human behavior and 
the underlying subjective experience of consciousness, with its bewildering repertoire of possible contents. However, the proposal that the same mechanisms underlying physical complexity also underlie 
the biological complexity of the brain is surprisingly recent. The description of the dynamical rules governing neurons at the
microscopic level \cite{hh} and the first mathematical demonstration of a second order phase transition in a many body system \cite{onsager} were almost contemporaries. However, 
these developments were separated by a large temporal gap from the first proposals of criticality in brain dynamics \cite{bak,Chialvo99}. Being a relatively recent proposal, the consequences of
such hypothesis are still far from clear. In the present essay we first discuss recent empirical evidence favoring this hypothesis, focusing on the presence of a second order phase transition in large scale brain dynamics \cite{enzo2012}, and then explore some possible consequences.

\section{Scale invariance in brain dynamics: Early findings}
 
The early evidences supporting criticality as a plausible dynamical regime for brain activity can be roughly classified according to the spatial and temporal 
resolutions at which they were obtained from the  experimental techniques used for the recording of brain activity. 
The results exposed in this section are important to introduce the recent finding of a second order transition in large scale brain dynamics, since the strategy used
for its uncovering can be regarded as a transfer of analysis techniques used at microscopic scales to the macroscopic large scale domain.

A landmark of the critical regime which occurs during a second order phase transition is the divergence of correlation length. As order emerges, the constituents of the system must organize themselves
instantaneously. Also at this point, any external perturbation will also have the highest impact on the system, as measured by the susceptibility. For that to occur, the dynamics of individual units must be mutually influenced even for those which are macroscopically separated and not directly connected. 
The divergence of correlation length 
implies scale invariance (i.e., fractality) in the system, as the presence of a characteristic scale would violate the divergence required by the instantaneous onset of the ordered phase.
In fMRI experiments, it has been demonstrated that functional connectivity networks are scale invariant \cite{Eguiluz} and, most remarkably, are virtually indistinguishable
from correlation networks obtained in the Ising model at the onset of the second order phase transition \cite{Fraiman}. Also, this scale invariance has been
directly demonstrated for fMRI data \cite{expert}, as well as the divergence in the correlation length \cite{fraiman2010}. 

We emphasize that this evidence of scale free brain dynamics at the
large scale domain has insofar treated brain activity as a continuously variable. Whether this was the case or its continuous nature emerged
as an artifact due to experimental and physiological blurring remained unknown.
On the other hand, experiments at smaller (microscopic scales) have concentrated in the description of brain activity as discrete avalanching events, spreading
throughout the cortex in a scale free fashion \cite{plenz}. These scale free avalanches have been exposed using electrophysiological techniques in different settings \cite{enzo2012, peterman, sidarta}. 
Self organized critical systems are known to dissipate energy in form of power law distributed avalanches \cite{Bak87, bak} , hence, this is direct evidence favoring the hypothesis that brain
 achieves critical properties through self-regulation and does not require fine tuning of parameters. We re-emphasize that experiments at this scale, either single unit or Local Field Potential (LFP) recordings, demonstrate scale-free intermittence and bursting, with
the intensity of the discrete burst obeying a power-law distribution.
 \begin{figure}[htb]
\centering{\includegraphics[width=0.75\textwidth,clip=true]{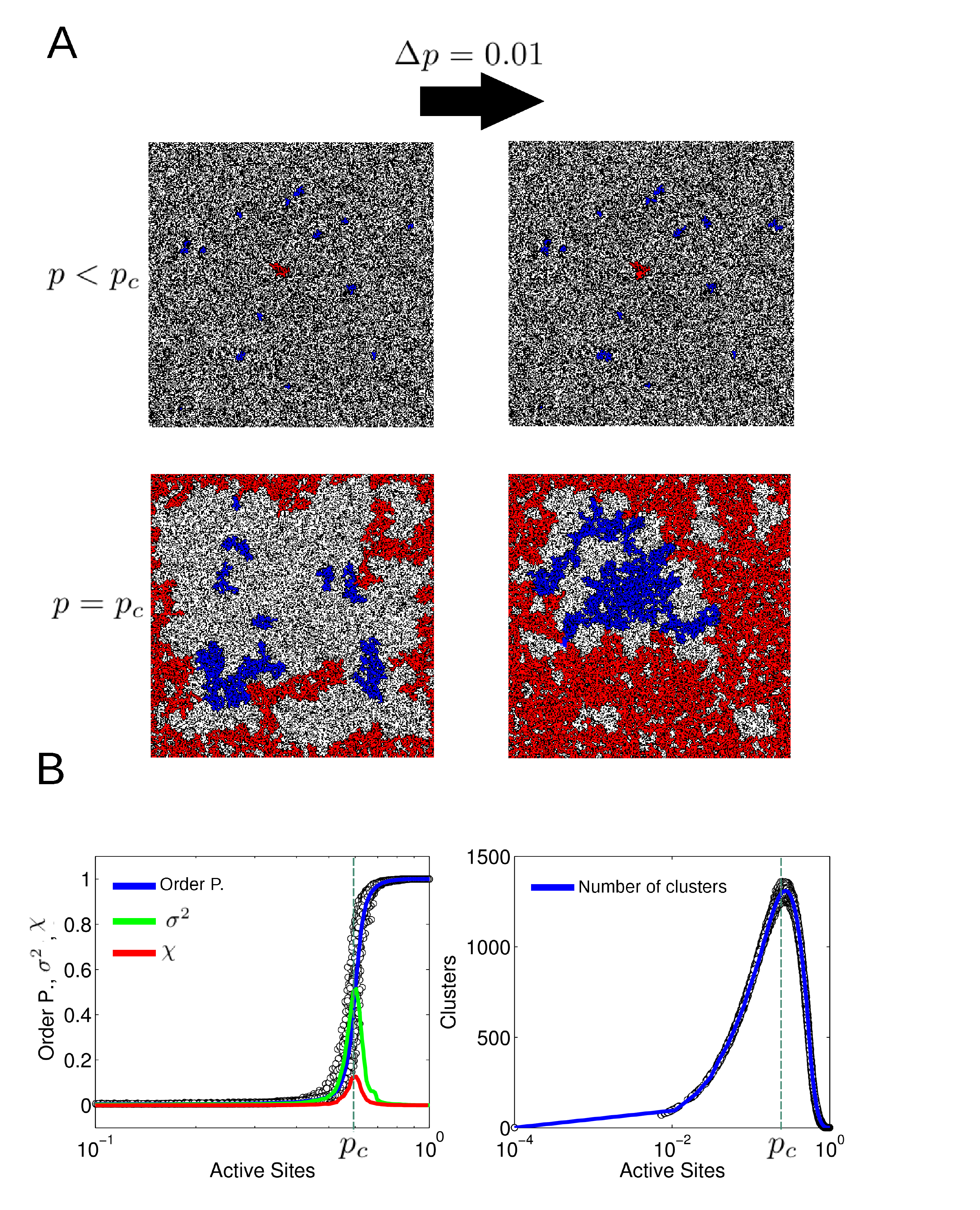}}
\caption{Order, disorder and susceptibilty in a second order percolation phase transition. A. The state of the system (black dots represent unnocuppied 
sites, white represents occupied sites) at sub-critical and critical concentrations. The effect of a small concentration change on the largest cluster (order parameter, red cluster)
and other smaller clusters (blue) is shown, demonstrating how a small perturbation can induce large changes near the critical point (but not
during the sub-critical phase) B. Order parameter, its variability ($\sigma^2$) and susceptibility ($\chi$) as a function of the control
parameter (left) and percolation diagram of active sites vs. number of clusters (right). In this case, the control parameter is defined as the ratio
of occupied to the total number of sites (concentration of active sites).}
\label{fig1}
\end{figure}
\section{A phase transition in cortical dynamic and its consequences}

As compelling as the above  experimental findings of scale invariance may be, they fail to reveal clues about its origin. 
Eventually, what is needed is to identify 
from the data the fundamental elements of a second order phase transition, namely the dynamical changes in an order parameter as a function of some control parameter. The 
derivation of those is not required solely for  theoretical reasons but is also  of important practical relevance: a control parameter allows to
quantify the ``degree of criticality'' present in the system. As in any finite non-equilibrium system, brain activity fluctuations impose 
spontaneous changes in the dynamical regime. An order and control parameter would allow, for instance, to dissipate the idea that the dynamical regime of the
brain is fixed and therefore to study the impact of criticality in the spontaneously fluctuating aspects of behavior and cognition.

A leap forward in this direction provided for the first time a way to estimate  these parameters \cite{enzo2012}. The analysis was carried out at the level of whole-brain
human fMRI data, an imaging modality known for its excellent spatial resolution. The control parameter defined here is roughly equivalent to the instantaneous
global activity level. The order parameter is equivalent to the size of the largest cluster of activated cortical activity (normalized by
the global activity level). With these definitions, a scatter plot of order vs. control parameters showed a sharp transition around a critical level of global activity,
yielding a diagram which resembles those derived for other systems undergoing a second order phase transition.  

Before exposing the rationale behind this
 choice of order and control parameters, we will briefly review the key insights which allowed their identification. As anticipated in the previous section,
these are related to the fusion of analysis techniques used at different scales, allowing a unified interpretation of the dynamical properties of the brain.

The most important insight of this work was its departure from techniques which estimate time averages of activity (during extended periods of time) and to focus on charactering spatiotemporal instantaneous activity. 
Since the early days of fMRI, the Blood Oxygen Level Dependent
(BOLD) signal is treated as a continuous smooth signal, even in spite of a large body of evidence showing that neural activity (at all scales) happens in bursts. Following this natural
line of thought, we discarded low excursions of the hemodynamic signals and focused on large amplitude events \cite{enzo2}. This procedure is analogous to the LFP thresholding used at
smaller scales to uncover avalanching activity \cite{plenz,plenz3}.
In formal terms, this is achieved by the reduction of the signal to a point process, which in turn is constructed by the introduction of a Poincar\'e section of the BOLD signal, as usually done in dynamical systems \cite{poincare1, poincare2, poincare3, poincare4, poincare5, poincare6}.
Notably, the resulting point-process remarkably resembled the results of resting-state BOLD signal deconvolution, giving formal support for its introduction.

The point process allows, for the first time, the visualization {\it at each time step} of the brain spatial pattern of activation. Thus, this approach allows us to perform a true spatiotemporal analysis of fMRI data.
 This information is then used to identify both parameters:
the control parameter, which is defined as the sum of all voxels above a threshold (i.e. those active in the point process) and the order parameter
as the size of the largest cortical cluster. The computation of these parameters from the fMRI data is  straightforward  applying a cluster labeling algorithm  to the spatiotemporal point process.

These definitions for order and control parameters are of clear interpretability in terms of degrees of order/disorder. 
Consider a moment in time in which the brain has a low value of the order parameter, i.e. the largest cluster of cortical activation is small compared to the total activated gray matter. It is possible to accommodate a vast number of  such clusters on the cortical surface, hence, the entropy is high, as corresponds to a disordered state. Following
the increase of the control parameter a point is reached in which a giant cluster of cortical activity emerges, comprising and integrating many different functional systems. 
Since all activations have coalesced
into this single large cluster, there is little room for variation: entropy is lower, as corresponds to an ordered state. In the extreme, a single cluster spans all active voxel, giving only one possible state (a globally activated state with zero entropy).

It is useful to view  these parameters in analogy to those used to quantify the dynamical regime of the percolation model, 
in which occupied (active) sites are placed on a lattice with different
concentrations (defined as the ratio of occupied to total sites) and the size of the largest cluster is taken as a measure of order.
 Figure \ref{fig1} shows an illustration of the model and the evolution of different quantities due to changes in the control parameter, which in this case is the concentration of sites. 
Despite the fact that the percolation model has no dynamics, it serves an an useful example to introduce the ideas of order parameter, control parameter and the maximum variability and susceptibility of the critical state, as discussed below.
\begin{figure}[h]
\centering{\includegraphics[width=.9\textwidth,clip=true]{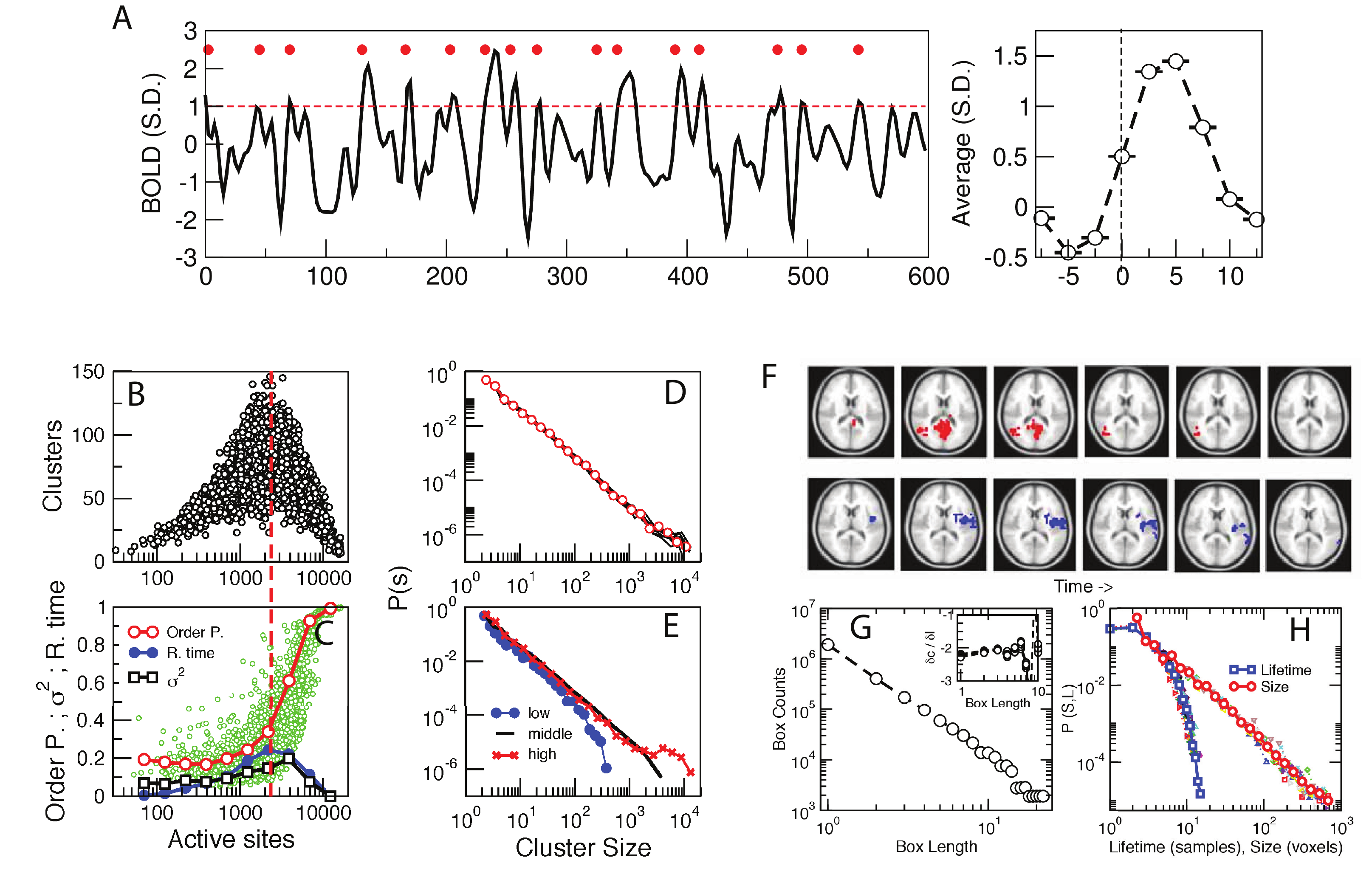}}
\caption{Identification of a second order phase transition in fMRI dynamics. A. Decomposition of an hemodynamic signal into a point process (left) which
generates an stereotypical waveform similar to the hemodynamic response function, the fMRI signal response to a single and discrete electrophysiological activation (right). 
B. Percolation diagram of active sites vs. number of clusters. C. Order 
parameter, variability and residence times as a function of the control parameter (active sites). D. Distribution of cluster sizes. E. Distribution of 
cluster sizes at the three different dynamical regimes. F. Examples of avalanches triggered from the visual (up) and insular (bottom) cortex. G. Fractal
dimension of the active sites obtained using a box counting algorithm. H. Distribution of avalanche sizes and avalanche lifetimes. Figure redrawn from \cite{enzo2012}.}
\label{fig2}
\end{figure}
In Fig. \ref{fig2} we summarize the evidence of a second order phase transition found recently \cite{enzo2012}. We introduce the point process decomposition of the
hemodynamic signal, we show the evolution of the order parameter vs. the control parameter, as well as the variability and the residence times (the definition
is introduced below), as was similarly presented in Fig. \ref{fig1} for the percolation model. This figure also presents results on the dynamical 
evolution of clusters, which spread as avalanches, a result consistent with previous findings at smaller scales \cite{plenz}.

Clusters (or ``blobs'', as they are usually refereed to in the neuroimaging community) represent co-activations (as evidenced by the point-process)
of contiguous brain regions which are usually associated with sensory, motor, or higher order cognitive systems. For instance, statistics over time averages
in a visual stimulation experiment would reveal a blob in the visual cortex (and similarly in other areas for other experiments). Seen this way, clusters represent a section 
or (level set) of the Statistical Parametric Map (SPM) which encodes the spatial distribution of the model fit statistical significance. Such model fit is performed during
extended periods of time. However, the point process not only reveals
that this activations appear spontaneously in the resting state but also shows their relationship with the dynamical regime: order means a shared co-activation
(integration) of the processes underlying these clusters, disorder means an increased segregation. In between, the transition point arguably represents
an optimum balance of segregation/integration. It also corresponds to the point in which the brain displays the higher variability in its repertoire of 
states, as evidenced by a peak in the variability of the order parameter vs. the control parameter (Fig. \ref{fig2}B, Fig. \ref{fig2}C).

As noted above, the brain fluctuates in and out of this regime. However, we demonstrated that 
most of the time it stays at the transition point, as evidenced by the computation of residence times (Fig. \ref{fig2}C. Residence times quantify how much time the system
spends at each possible state of the space of dynamical variables. In this case, we have tracked the state of the system using the order and control parameters and
we demonstrated that the system stays for more prolonged periods at the state corresponding to the critical control parameter. Incidentally, this may explain why direct evidence
for the presence of a second order transition had to wait so long: since on average the brain spends most of the time in the transition point, any approach
which is based on time averaged quantities is bound to highlight only the critical state, but not the super- (disordered) and sub- (ordered) critical states, as can be done
with the point process approach.

Quantification of the cluster spatio-temporal evolution demonstrates scale-free avalanches spreading in fractal-like structures (with dimension
slightly greater than two, see Fig. \ref{fig2}F, Fig. \ref{fig2}G, Fig. \ref{fig2}H). This result is a confirmation of what was previously known for brain dynamics at smaller scalers by means of electrophysiological
experiments \cite{plenz}. However, it highlights the fact that predictions based on properties of the critical state are manifest in all spatial and temporal scales 
of brain dynamics. Scale invariance dictates that avalanches of activity must follow the same distribution regardless of their size and that events as large
as the size of the system must be found. Both predictions are clearly confirmed by analysis of fMRI dynamics in the  computation of power law exponents. 
Importantly, the theory of phase transitions provides a set of critical exponents which may be computed for this data and shed light on a possible
universality class for brain dynamics.

Finally, a peak in variability is related to a peak in the susceptibility of the system near the critical point of a  second order phase transition. To introduce the concept of susceptibility, imagine a physical system upon which we exert forces. We expect such forces to elicit changes in the system. Susceptibility can be roughly defined as the ratio of the elicited response and the exerted influence. In finite critical systems, susceptibility is maximized at the critical point, therefore, the capacity of the system to react to external changes is maximized. Given the need of flexibility and reactiveness of the 
brain, the endowment of maximum susceptibility due to the critical state is a very attractive possibility from an evolutionary point of view. One must contrast this non-equilibrium view with that of many models (such as the attractor networks of the Hopfield model \cite{hopfield}) in which, after the system reaches its final state, it has a 
vanishing susceptibility (since small changes which could elicit large responses at the critical state are not capable of shifting the system from the equilibrium point or attractor).
The relationship between equilibrium and non-equilibrium models and the inclusion of noise in dynamical equations will be discussed in the next session.
 \begin{figure}
\centering{\includegraphics[width=.95\textwidth,clip=true]{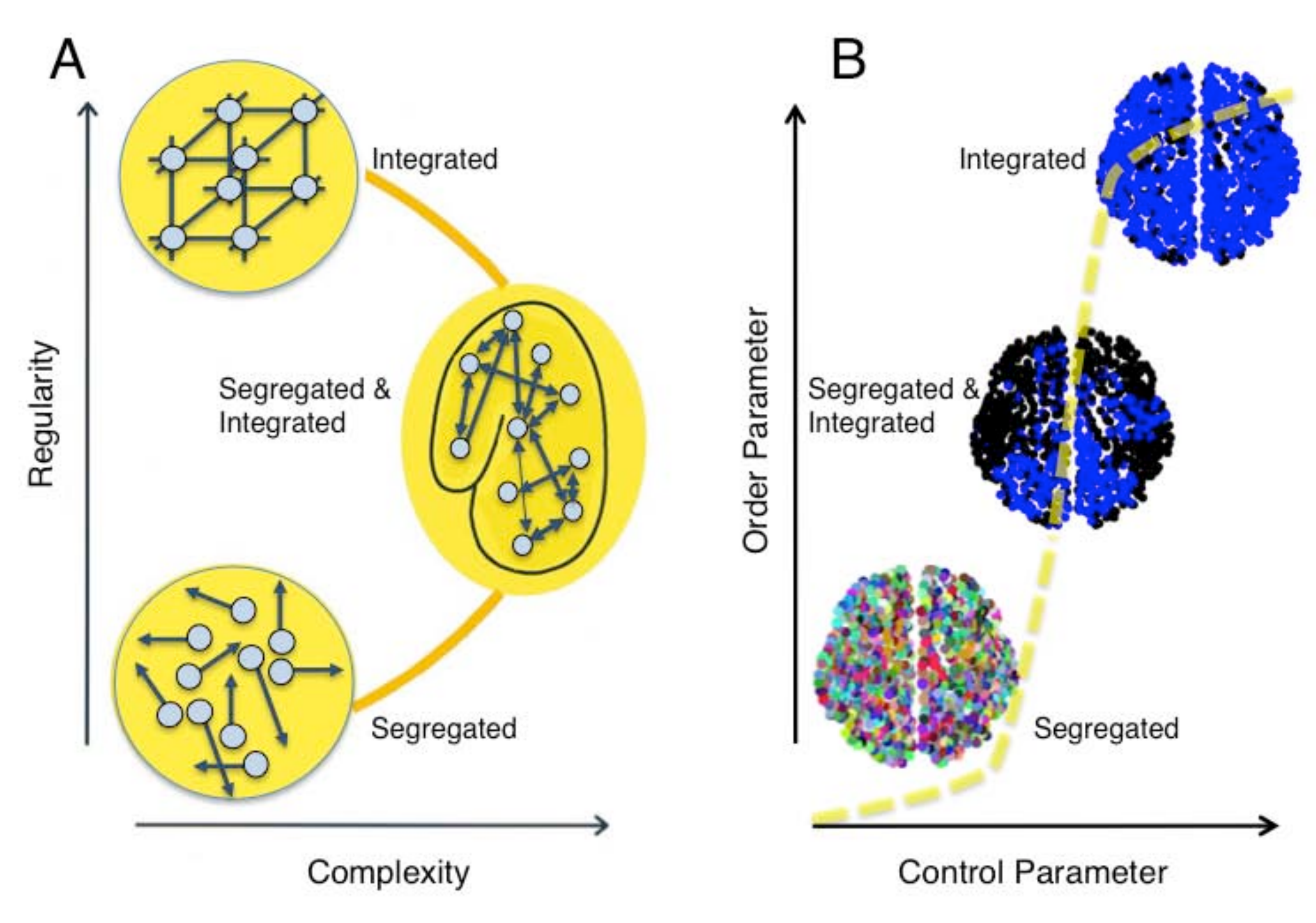}}
\caption{Two views of the integration-segregation dilemma.  Panel A shows a \emph{structural} point of view, in which neural structures exhibit different regularity and complexity. According to this, the structural connectivity of  the human brain has a high complexity and intermediate regularity, with a crucial balance of integration and segregation acquired trough evolutionary selection \cite{tononi3}. Panel B depicts a \emph {dynamical} alternative. According with the proposal discussed here,  the three type of regimes in A (with higher or lower complexity)  can be dynamically (and transiently) generated by any system undergoing a second order phase transition (even with trivial and homogeneous structural connectivity), because they represent  the three generic dynamical regimes of the system. Colors in the three graphs label the clusters with coherent activity, thus in the most ordered regime (top right) the entire brain is active and coalesces into a single activated cortical cluster, whereas in the most disordered one (bottom left), each brain region acts independently. It is only at criticality (middle) that coherent clusters of all sizes are possible, thus optimizing the integration/segregation balance. Panel A  redrawn from the figure in Box 2 of \cite{tononi3} rotated 90 degrees counterclockwise.  Panel B redrawn from \cite{ariel}.}
\label{fig3}
\end{figure}
\section{Order/disorder vis a vis integration/segregation} 

As discussed already elsewhere \cite{chialvo2010} the idea of a continuous phase transition in brain dynamics is related to the ability of the brain
 to simultaneously integrate and segregate information, a point championed by Edelman, Tononi and Sporns \cite{tononi1,tononi, tononi2, tononi3}. 
In their own words: \emph{``Nervous systems facing complex environments have to balance two seemingly opposing requirements. First, there is a need
 quickly and reliably to extract important features from sensory inputs. This is accomplished by functionally segregated (specialized) sets of neurons,
 e.g. those found in different cortical areas. Second, there is a need to generate coherent perceptual and cognitive states allowing an organism to 
respond to objects and events, which represent conjunctions of numerous individual features. This need is accomplished by functional integration of the activity of specialized neurons through their dynamic interactions"}  \cite{tononi1}.

The cartoons in Fig.\ref{fig3} show schematically that low complexity is expected for interactions occurring  at both extreme degrees of order and 
disorder. It is only at the intermediate level that complexity peaks, when diverse mixtures of order and disorder are present.  
Panel A (taken from \cite{tononi3}) illustrates the point for the structural case, in which the brain structure molded trough evolution is able to
 optimize both integration and segregation.  The case in panel B represents  the dynamical scenario (taken from \cite{ariel}) of a second order 
phase transition. In this case, as the control parameter is increased, the order parameter also increases, slowly first and then suddenly  at the 
transition point. The overimposed pictures represent the top view of the brain from  simulations of its activity at different regimes 
(unpublished data from \cite{ariel}). The three examples are the type of clusters 
(as defined already for the results in Fig \ref{fig2}) found at disordered, intermediate and ordered regimes. 
It is only at the critical regime that clusters of all sizes can be observed, which in dynamical terms represent the balance between high 
integration (a few big clusters) and high segregation (many small clusters). 
Indeed this mixture is characterized by a scale invariant distribution of cluster sizes (see Fig. \ref{fig2}D for the distribution found experimentally).

Before moving away from Fig \ref{fig3}, it need to be noted that, although unlikely to be a realistic possibility, here and only in terms of considering mechanisms, 
the complexity exhibited by the networks depicted in Panel A  could also arise from a phase transition. 
An example can be bond percolation, for which is known that the most complex network structures arise at the percolation threshold.

It is important to note two additional points  about  dynamics and structural complexity. First of all, we do not stress the segregation/integration 
 balance by itself: it is considered simply a consequence of the physics of phase transitions, which are, arguably, of a more fundamental nature. 
Still, the relation between these concepts clearly deserves further theoretical and experimental investigations. 

The second point to remark is a bit subtler: criticality endows an arbitrary graph with similar dynamic complexity 
(i.e., similar integration/segregation balance) than the structural complexity of the brain. However, this is not meant to imply that any 
arbitrary structure, provided with critical dynamics, can ``think'' like a brain. An understanding of the relation between the connectome 
(an exhaustive description of all \emph{structural} connections) and what we call cognitome (an equally exhaustive description of the \emph{functional} repertoire of the brain)
 and the role of dynamics is still ahead and much work still needs to be done.
 
A lesson taught by the findings on critical phenomena in brain dynamics is that the explanative power of physical theories should not be disregarded
 in biology just because it provides all encompassing, holistic explanations as opposed to detailed, reductionistic descriptions of myriads of
 experimental facts. A glaring example is the question of what role noise plays in brain dynamics and where it is generated. 
Often theories postulate a ``key role''  for noise to adequately explain response variability in healthy brain function and proceed to explain the
  observations in the light of very detailed noisy processes.  In the same direction, the information content of the brain BOLD signal's variability
 {\it per se} received increasing interest. For instance,  it was shown recently \cite{noise} in  a group of subjects of different age, 
that the BOLD signal variability (standard deviation) is a better predictor of the subject age than the average.  
Furthermore, additional work focused on the relation between the fMRI signal variability and task performance, and concluded that faster and
 more consistent performers exhibit significantly higher brain variability across tasks than the poorer performing subjects \cite{Garrett}.
From  the current perspective, all these observations have the same underlying explanation: maximum variability is a straightforward consequence 
of the critical regime as shown recently \cite{fraiman2010}.  Noise is endogenous to non-equilibrium systems at the transition point and it 
needs not to be introduced as an ad-hoc equilibrium explanation every time a neurobiological fact displays a large degree of variability.

\section{Summary and outlook}

In this essay we have exposed recent experimental evidence demonstrating the presence of a second order phase transition in human brain dynamics, 
we have explored some of its consequences. In doing so, we have stressed that the  application of these concepts should be aimed as a unifying physical
 explanation of the brain. We have also stressed that biological theories should take advantage of the all-encompassing framework provided by physical
 theories (in particular, that of critical phenomena) instead of relying on reductionistic ad-hoc explanations tailored for each experimental fact.

After establishing that resting state brain activity displays many signatures of sub, super and critical states, a question clearly needs to be addressed: 
what is the neurobiological role of these dynamical regimes? Since the brain continuously enters and leaves the critical regime (but remains most of the time at this point),
one could be tempted to speculate about the possibility of a more permanent displacement. In other words, if criticality is important for healthy
brain function, what happens if this property is lost? To gain insights on this question we propose to study brain states which radically differ
from wakeful rest (by far the most studied condition in fMRI resting state analyses). Examples could be deep sleep, anesthesia, coma, as well as different states
of consciousness induced by drug intake. We believe that new exciting venues of research will be open to clarify the role of these fluctuations around criticality  in relation to these different neurobiological states.
\\

\begin{theacknowledgments}
Work supported by NIH (USA), CONICET (Argentina) and the MCyT (Spain). ET was partially funded by the Bundesministerium fur Bildung und Forschung (grant 01 EV 0703) and LOEWE Neuronale Koordination Forschungsschwerpunkt Frankfurt (NeFF). 
\end{theacknowledgments}

\bibliographystyle{aipproc}

\end{document}